\documentstyle[prl,aps,multicol,graphicx]{revtex}

\begin{document}
\title{Systematics of two-component superconductivity in $YBa_{2}Cu_{3}O_{6.95}$
from microwave measurements of high quality single crystals}
\author{H.Srikanth, Z.Zhai and S.Sridhar}
\address{Department of Physics, Northeastern University, 360 Huntington Avenue,\\
Boston, MA 02115}
\author{A. Erb and E. Walker}
\address{DPMC, Universite de Geneve, Geneve, Switzerland}
\date{\today}
\maketitle

\begin{abstract}
A systematic set of microwave measurements of the surface impedance ($%
Z_{s}=R_{s}+iX_{s}$) of $YBa_{2}Cu_{3}O_{6.95}$ single crystals (called $%
YBCO/BZO$) grown in $BaZrO_{3}$ crucibles reveal new properties that are not
directly seen in similar measurements on other $YBCO$ samples. The complex
conductivity $\sigma =\sigma _{1}-i\sigma _{2}$ obtained as a function of
temperature ($T$) from the surface impedance data shows two key features :
(1) A new conductivity peak in $\sigma _{1}(T)$ around $80K$ in addition to
peaks at $30K$ and $92K$, and, (2) extra pairing below $65K$ in addition to
the onset of pairing below the bulk $T_{c}$ of $93.4K$ as revealed in $%
\sigma _{2}(T)$. These features are present in all 3 $YBCO/BZO$ crystals
measured and are absent in $YBCO$ crystals grown by other methods. These
results show that in addition to pairing at $T_{c}=$ $93.4K$, an additional
pairing channel opens up at $(\,\sim \,65K)$.

High pressure oxygenation of one of the crystals still yields the same
results, and shows that the data cannot be due to unwanted macroscopic
segregation of $O$ -deficient regions. Systematics on three single crystals
show that the height of the quasiparticle conductivity peak at $80K$ in the
superconducting state is correlated with the inelastic scattering rate in
the normal state. Close to $T_{c}$, $\sigma _{2}(T)\sim (T_{c}-T)$,
indicating a mean-field behavior and inconsistent with $3DXY$ fluctuations.

{\em A single complex order parameter cannot describe these data, and the
results suggest that at least two superconducting components with
corresponding pairing temperature scales (}$T_{A}\sim 65K${\em \ and }$%
T_{B}=T_{c}=93.4K${\em \ ) are required}. Comparison to model calculations
considering various decoupled two-component scenarios ( $A+B=\,d+s,s+d,d+d$)
are presented. The comparison shows that the experimental data do not
distinguish between these various scenarios, however the data do require
that one of the components be an order parameter (OP) with nodes in the gap,
such as a $d-$ wave OP. Fit parameters to the calculations using the
different scenarios are presented. These components are naturally present in
all $YBCO$ samples, however impurities appear to suppress the pair density
of the low temperature $A$ component and lead to greatly enhanced scattering
of the high temperature $B$ component. Overall, our results strongly suggest
the presence of multiple pairing temperature scales and energies in $%
YBa_{2}Cu_{3}O_{6}._{95}$.
\end{abstract}

\begin{multicols}{2}

\section{Introduction}

The order parameter of high temperature superconductors has been extensively
studied recently, and a consensus seems to have emerged in favor of a d-wave
order parameter (OP) \cite{VanHarlingen95}. However there are some notable
indications (see ref. \cite{Muller95,Koltenbah97} for a summary) which
suggest that, particularly in the most widely studied material, $%
YBa_{2}Cu_{3}O_{6.95}$, a pure d-wave OP may not occur and that there are
indications of a multi-component OP ({\em eg.} s + d).

Material purity is crucial for studies of fundamental physical properties
not obscured by impurity-related artifacts. This fact has been validated
time and again in experiments on oxide superconductors as well as other
materials. It is now accepted that improvement in material quality has often
resulted in a better understanding of the physical properties in solid state
systems.

The recent growth of $YBa_2Cu_3O_{7-\delta }$ single crystals in $%
BaZrO_3(BZO)$ crucibles has ushered in a new generation of ultra-pure
samples \cite{Erb95a}. This growth method avoids the critical problem of
crucible corrosion and leads to single crystals with extremely clean
surfaces and purity exceeding $99.995\%$. In contrast, the best crystals
grown in conventional crucibles like $Au$ and yttria-stabilized-zirconia $%
(YSZ)$ have final reported purities of $99.5-99.95\%$ \cite
{Casalta96,Ikuta96} A\ number of experiments on $YBCO/BZO$ crystals have
revealed new features which are either greatly suppressed or not present in $%
YBCO/YSZ$ samples. We have recently observed novel features in the microwave
surface impedance of $YBCO/BZO$ which clearly indicates the presence of two
superconducting components \cite{Srikanth97a}.

In this paper, we present further evidence on the systematics in three $%
YBCO/BZO$ crystals. The results obtained confirm our earlier findings \cite
{Srikanth97a}, {\em viz.} : (1) A new conductivity peak in $\sigma _{1}(T)$
around 80K which is correlated with the normal state scattering rate and (2)
Extra pairing below 65K in addition to onset of pairing below the bulk $%
T_{c} $ of $93.4K$ as revealed in $\sigma _{2}(T)$. {\em A single complex
order parameter cannot describe these data and the data indicate the
presence of at least two pairing processes, leading to at least two OP
components}. Model calculations considering various simple two-component
scenarios (decoupled $d+s$ and $d+d$) and are presented. We show that the
two-component picture can provide a quantitative description of the data,
whereas a single order parameter (of any symmetry) does not. In addition the
two-component picture can provide a description of the data on the earlier
generation of $YBCO/YSZ$ crystals also. Analysis of the data close to $T_{c}$
is not consistent with $3DXY$ fluctuations and displays a mean-field
behavior. Systematics on three single crystals show a correlation between
the inelastic scattering rate in the normal state with the quasiparticle
conductivity peak at 80K in the superconducting state. We show that while a $%
T$-dependent scattering rate is required to quantitatively describe the
conductivity peaks, the two-component scenario provides a natural
explanation for the location of the $30K$ and $80K$ peaks in $\sigma _{1}(T)$
as arising from pairing at $65K$ and $93K$ respectively, in contrast with
earlier interpretation of $YBCO/YSZ$ data which associated the $30K$ peak
with pairing at $93K$. Overall, our results show that {\em a pure $d-wave$
state does not occur in $YBa_{2}Cu_{3}O_{6.95}$} and strongly suggest the
presence of multiple pairing energies in $YBa_{2}Cu_{3}O_{6}._{95}$.We note
that several microscopic pairing scenarios are consistent with the
conclusions of this paper.

\subsection{Properties of $YBCO/BZO$ crystals}

One of the causes for the presence of impurities in $YBa_2Cu_3O_{7-\delta }$
single crystals is the random substitution at the $Cu$ chain sites by trace
amounts of crucible constituents like $Au$ during the melt growth process.
This results in local variation of oxygen vacancy distribution, introduction
of magnetic moments and other local defects \cite{Erb96a,Stoney,Becht}.
Although the overall $T_c$ and sharpness of the superconducting transition
may not be affected by a combination of these elements associated with
impurities, important features of the superconducting ground state like the
order parameter symmetry, scattering, superfluid density etc. are likely to
be influenced. It is precisely these local impurities which are eliminated
in $(BZO)$ grown $YBa_2Cu_3O_{7-\delta }(YBCO)$ crystals thus providing an
opportunity to probe the intrinsic ground state properties free from
defects. Elimination of the metallic impurities also leaves oxygen
stoichiometry as the only variable which needs to be controlled \cite{Erb96a}%
.

It is important to emphasize at this juncture that several new results have
been obtained on these new generation crystals by a number of experimenters
using a variety of probes like thermal, magnetic and electrodynamic
response, and have led to a clearer picture of the nature of
superconductivity in $YBa_{2}Cu_{3}O_{6.95}$. These results are :

(1) Clear imaging of the flux lattice of $YBCO$ using low temperature STM 
\cite{MaggioAprile95}. Thus far this has only been feasible with the new $%
YBCO/BZO$ crystals.

(2) ``Fishtail effect'' in magnetization {\em eliminated} in high pressure
oxygen annealed $YBCO_{7.0}/BZO$ and optimally doped $YBCO/BZO$ samples \cite{Erb3}.

(3) Schottky contribution to specific heat {\em suppressed }in{\em \ }$%
YBCO_{7.0}/BZO$ indicating the total absence of magnetic moments \cite
{JYGenoud96,Roulin96}.

(4) The microwave conductivities $\sigma _1(T)$ and $\sigma _2(T)$ exhibit 
{\em two} distinct features not consistent with a single superconducting
order parameter\cite{Srikanth97a}.

(5) The vortex lattice imaged with STM shows two different regions which is
either representative of two superconducting components or two types of
ordered oxygen clusters \cite{MaggioAprile96}.

(6) Clear observation of a direct first order melting transition in $%
YBCO_{7.0}/BZO$ from a vortex lattice to liquid without an intervening
glassy state \cite{Indenbom97}.

(7) Evidence of extremely low $J_{c}$ in $YBCO_{7.0}/BZO$ crystals
indicative of very low pinning.

In an attempt to bring forth the essential differences, we have presented a
comparison of some material and physical properties of $YBCO/YSZ$ and $%
YBCO/BZO$ crystals in a tabular form in Table I.

\section{Experimental Results}

The microwave measurements were carried out in a $10GHz$ $Nb$ cavity using a
``hot finger'' technique \cite{SSridhar88}. We measure the surface impedance 
$Z_{s}=R_{s}+iX_{s}$ and penetration depth $\lambda (=X_{s}/\mu _{0}\omega )$
as functions of temperature $T$, from which we extract the complex
conductivity $\sigma _{s}=\sigma _{1}-i\sigma _{2}$. This method has been
extensively validated by a variety of measurements on cuprates and
borocarbide superconductors \cite{TJacobs95,TJacobs95d}.

Three single crystals (labeled AE103, AE105 and AE180 and typically $%
1.3\times 1.3\times 0.1mm^{3}$ in size) from different batches grown in $BZO$
crucible and one crystal (labeled AEXX) of comparable dimensions grown in $%
YSZ$ crucible were measured to study the systematics in the different
samples. While standard oxygen annealing procedures at atmospheric pressure
were followed to obtain optimally doped crystals with oxygen stoichiometry
around $O_{6.95}$ in AE103, AE105 and AEXX, the AE180 sample was annealed at
a pressure of $\ 100$ $bars$ for 20 hours. Note that this high pressure
annealing results in elimination of the formation of oxygen vacancy clusters
and as a consequence, the ``fishtail'' anomaly in magnetization is
suppressed. The crystals (AE103, AE105 and AE180) grown in $BZO$ had $%
T_{c}=93.4K$ and AEXX had $T_{c}=92.4K$. All four crystals exhibited very
narrow transitions in $R_{s}(T)$ at 10 GHz of $<0.3K.$

The temperature dependence of surface resistance $R_{s}(T)$ and change in
penetration depth $\Delta \lambda (T)$ were presented for AE103 and AEXX
crystals and discussed in detail in an earlier publication \cite{Srikanth97a}%
. In this paper, we show that the data for AE105 and AE180 crystals also
show the new features seen in the AE103 single crystal. The data for all the
three high purity $YBCO/BZO$ crystals clearly reveal a new bump in the
vicinity of $60K$ which is not seen in $YBCO/YSZ$. Also the estimated London
penetration depth $\lambda (0)$ for $YBCO/BZO$ samples are $\simeq 1000\AA $
which is lower than the value of $1400\AA $ estimated for the $YBCO/YSZ$
crystal. We obtain the absolute value of $X_{s}$ by equating $R_{n}=X_{n}$
above $T_{c}$.

\begin{figure}[tbph]
\begin{center}
  \includegraphics*[width=0.45\textwidth]{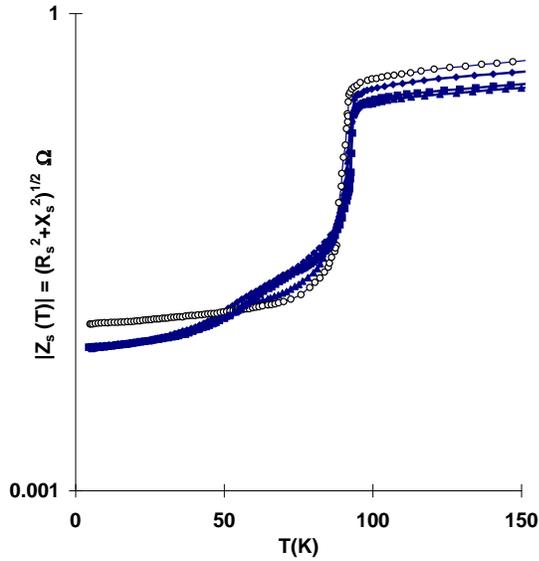} 
\end{center}
\caption{Magnitude of surface impedance amplitude $\mid Z_s(T)\mid $ of $%
YBCO/BZO$ and $YBCO/YSZ$ single crystals. Filled symbols are used for $BZO$
grown crystals: AE103 (filled squares), AE180 (filled triangles) and AE105
(filled diamonds). The data for $YSZ$ grown AEXX (open circles) is also
shown. The plot emphasizes the fact that $\lambda(0)$ for all the $YBCO/BZO$ crystals is
 lower than that of $YBCO/YSZ$. }
\label{amplitude}
\end{figure}

Since the absolute values of $R_s(T)$ and $X_s(T)$ are now known, we can
plot the amplitude $\mid Z_s(T)\mid \equiv \sqrt{R_s(T)^2+X_s(T)^2\text{ }}$%
as a function of $T$. Fig.\ref{amplitude} shows such a plot for the three $%
YBCO/BZO$ samples along with $YBCO/YSZ$ for comparison. The virtue of
plotting $\mid Z_s(T)\mid $ is that one can clearly see the lower $\lambda
(0)$ in $YBCO/BZO$ which effectively translates to a higher superfluid
density $(n_s/m)\sim 1/\lambda ^2$.

In the superconducting state, the complex surface impedance is related to
the complex conductivity and penetration depth: $Z_{s}(=R_{s}+iX_{s})=\sqrt{%
i\mu _{0}\omega /(\sigma _{1}-i\sigma _{2})}$. Since the absolute values of $%
R_{s}$ and $X_{s}$ are known, it is possible to extract the complex
conductivities:

\[
\sigma _1=\frac{2\mu _0\omega R_sX_s}{(2R_sX_s)^2+(R_s^2-X_s^2)^2} 
\]

\[
\sigma _{2}=\frac{\mu _{0}\omega (X_{s}^{2}-R_{s}^{2})}{%
(2R_{s}X_{s})^{2}+(R_{s}^{2}-X_{s}^{2})^{2}} 
\]

These quantities are important because they enable comparison with
microscopic theories. The pair conductivity $\sigma _{2}$ is a
representation of the superfluid density $\sigma _{2}=(\mu _{0}\omega
\lambda ^{2}(T))^{-1}$ and is a convenient way of probing how the condensate
builds up below the superconducting transition. The quasiparticle
conductivity $\sigma _{1}$ is determined by the quasiparticle density as
well as the quasiparticle scattering time.

\begin{figure}[tbph]
\begin{center}
  \includegraphics*[width=0.45\textwidth]{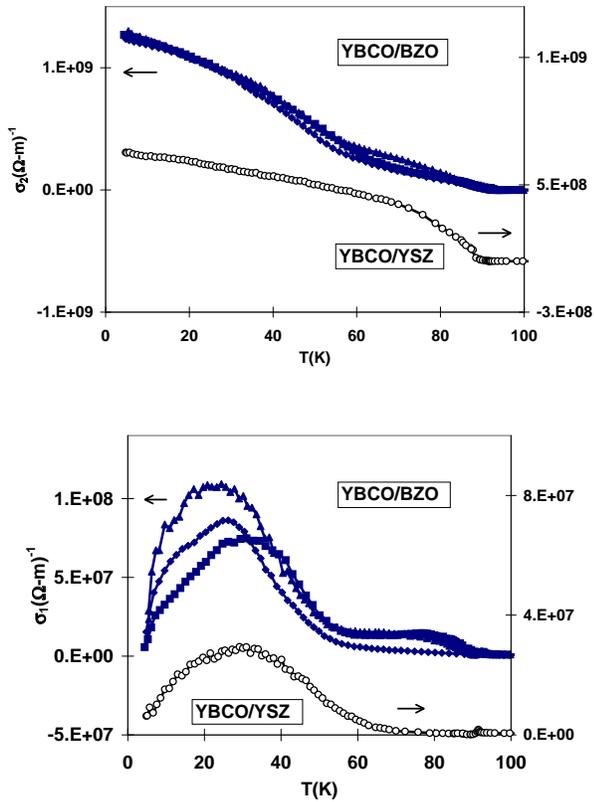} 
\end{center}
\caption{Complex conductivities $\sigma _2$ (top panel) and $\sigma _1$
(bottom panel). The $YBCO/BZO$ data are shown in filled symbols and the $%
YBCO/YSZ$ data in open circles. Note the extra pairing at around $60K$ in $\sigma_2$ 
and the new peak at $80K$ in $\sigma_1$ in $YBCO/BZO$, both absent in $YBCO/YSZ$ data. }
\label{sigmas}
\end{figure}

In Fig.\ref{sigmas}, the temperature dependence of $\sigma _1(T)$ and $%
\sigma _2(T)$ are plotted for all the four samples, AE103, AE105, AE180 and
AEXX. Two striking features emerge clearly from a comparison of the data:

(i) $\sigma _2(T)$ shows two distinct regions of variation above and below $%
\sim 60-70K$ in AE103, AE105 and AE180 samples in comparison with AEXX.
Given the fact that $\lambda (0)$ is lower in $YBCO/BZO$ this implies
enhanced pair conductivity in these samples.

(ii) The normal conductivity $\sigma _{1}(T)$ shows a new peak at around $%
80K(\sim 0.9T_{c})$ in $YBCO/BZO$ crystals which is absent in $YBCO/YSZ$.
This peak is greatly suppressed in AE105 as seen in Fig.\ref{sigmas}. We
show later that there is a correlation between this peak and the normal
state inelastic scattering rate of the samples. It is to be noted that the
normal conductivity peak at $\sim 30-35K$ and a very sharp peak near $T_{c}$
are present in all the samples.

\begin{figure}[tbph]
\begin{center}
  \includegraphics*[width=0.45\textwidth]{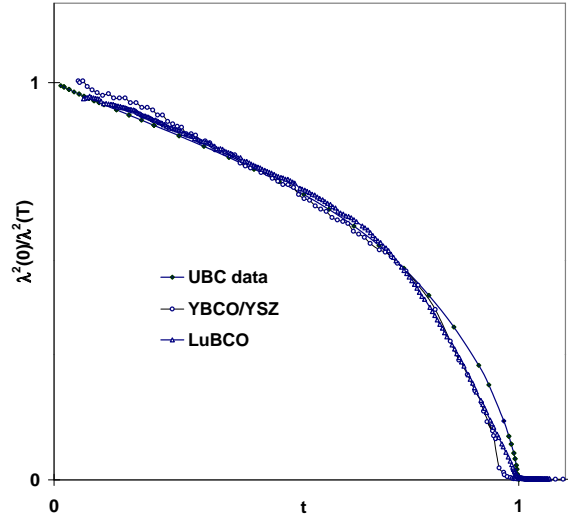} 
\end{center}
\caption{Superfluid density of the YBCO/YSZ single crystal (AEXX) plotted against reduced
temperature. Data for an untwinned $YSZ$ grown $LuBCO$ crystal and the UBC data are also shown for comparison.}
\label{n_s1}
\end{figure}

Note that these two new features are both absent in all previously measured
crystals. This is evident from the data on the $YBCO/YSZ$ sample shown in
Fig.\ref{sigmas}. We emphasize that the $YBCO/YSZ$ data is indeed
``canonical'' of the samples prepared by previous growth methods. This is
illustrated in Fig.\ref{n_s1} which compares the $YBCO/YSZ$ data with that
measured by us on an untwinned $LuBa_{2}Cu_{3}O_{6.95}$ sample \cite{Buan96}
and by the UBC group on $YBa_{2}Cu_{3}O_{6.95}$ \cite{Hardy93a}. Note the
excellent consistency of data on all previous samples with each other in Fig.%
\ref{n_s1}, and the systematic differences with the new data in Fig.\ref
{sigmas}.

\section{Analysis and Discussion}

A calculation of the high frequency conductivity requires proper
incorporation of scattering effects along with self-consistent calculations
of the gap and density of states. In addition anisotropy effects should also
be included for the cuprate superconductors. While this has been done \cite
{Klemm88,Hirschfeld94,Scharnberg97} for unconventional superconductors
including those with a d-wave OP, these approaches are not easily amenable
to comparison with experimental data.

Instead, in order to compare with the present experimental data we use a
simpler ``two-fluid'' model of the form \cite{Waldram97}:

\[
\sigma (\omega ,T)=\sigma _{1}-i\sigma _{2}=\frac{ne^{2}}{m}\left[ \frac{%
f_{n}}{i\omega +1/\tau (T)}-\frac{if_{s}}{\omega }\right] 
\]

where $f_{n}$ and $f_{s}$ represent the fractions of normal and superfluid
(with $f_{n}+f_{s}=1$), and $\tau $ is the relaxation time for normal
electrons. In this model, the normal electrons have damping with the usual
Drude conductivity at high frequencies, and the superconducting electrons
have inertia but no damping. The quasiparticle and pair conductivities can
be numerically calculated from:

\[
f_{s}=(1-f_{n})=1-2\left\langle \int_{0}^{\infty }\left( -\,\frac{\partial f%
}{\partial E}\right) d\epsilon \right\rangle 
\]

where $E=\sqrt{\epsilon ^{2}+\Delta ^{2}(\phi ,T)}$, and $\Delta (\phi ,T)$
is the gap parameter, and $<..>=\int_{0}^{2\pi }d\phi $ indicates an angular
average over $\phi $. The gap parameters are given by $\Delta (\phi
,T)=\Delta _{s}(T)$ and $\Delta (\theta ,T)=\Delta _{d}(T)\cos (2\phi )$ for
s-wave and d-wave superconductors respectively. For $\Delta _{s}(T)$ and $%
\Delta _{d}(T)$ we use appropriate mean-field temperature dependences. The
validity of this two-fluid approach, which is used frequently in s-wave
superconductors following Mattis \& Bardeen \cite{MattisBardeen58}, has been
discussed for d-wave superconductors by Hirschfeld, {\em et.al.} \cite
{Hirschfeld94}.

\subsection{Comparison with models}

We now discuss the comparison of the experimental data with detailed models
which are implemented assuming various conditions for the gap parameters.

\subsubsection{Single d-wave order parameter: Its failure to describe the
present results}

There is a general consensus that the superconducting order parameter
symmetry in cuprates is not a simple $s$-wave type as is the case in
conventional superconductors. A wealth of experimental results indicate
unconventional pairing and more specifically point to the presence of nodes
in the gap. A strong candidate which has been found to account for most of
the properties in cuprates is the theory of $d$-wave superconductivity. A
single complex $d$-wave order parameter has been proposed with a gap of the
form: $\Delta (T)=\Delta _{d}(T)cos2\phi $ where $\phi $ specifies the
orientation of the two-dimensional momentum of the Cooper pairs. This
expression leads to two important signatures, namely, the order parameter
goes to zero in certain $k$ directions and also changes sign as one goes
around the Fermi surface. Angle-resolved photoemission \cite{ZXShen93} and
phase sensitive SQUID experiments \cite{VanHarlingen95} have shown strong
evidence for this behavior. For experiments which measure quantities
averaged over $k$-space (like the microwave penetration depth), the
consequence of the $d$-wave gap will lead to features in the low temperature
dependence which should be consistent with the presence of a finite density
of states within the gap.

A linear low temperature dependence in YBCO single crystals first observed
by UBC group \cite{Hardy93a} (and subsequently confirmed by others including
ourselves in $YBCO$ \cite{TJacobs95,Mao95} and also $Bi:2212$ \cite
{TJacobs95d}) has been claimed as prime evidence for nodes in the gap, which
applies to $d$-wave symmetry among other possibilities. Rigorous
calculations of the microwave conductivity in the framework of $d-$wave
theories have also been done to explain the features observed in prior $%
YBCO/YSZ$ samples \cite{Hirschfeld94,Scharnberg97}.

In Fig.\ref{n_s1}, a plot of $(\lambda ^{2}(0)/\lambda ^{2}(T))$ vs. reduced
temperature $(t=T/T_{c})$ for $YBCO/YSZ$ single crystals is shown. A
comparison to detailed d-wave calculations has been carried out by us in an
earlier publication \cite{TJacobs95}. Comparison shows that although the low
temperature dependence is reproduced by calculations (not shown) using a
weak coupling $d-$wave order parameter, the agreement is not very good at
temperatures $T>0.5T_{c}$, suggesting that strong coupling effects are
needed. 

For the $YBCO/BZO$ single crystals, the low temperature penetration depth $%
\lambda (T)$ is indeed linear for all 3 samples measured up to $\sim 20-25K$
with a characteristic slope $d\lambda /dT$ of about $4.5\AA /K$, similar to
that of the $YBCO/YSZ$ samples. However deviation from linearity is observed
above $25K$ for the $YBCO/BZO$ crystals due to the onset of the broad $60K$
bump. This is clearly evident from the plots of $\sigma _{2}(T)$ presented
in Fig.\ref{sigmas} which shows a non-monotonic dependence for AE103, AE180
and AE105 samples when compared to the AEXX curve. While a
single order parameter may be reconciled at first glance for the $YBCO/YSZ$
data \cite{TJacobs95}, it is impossible to do so for the results on the $%
BaZrO_{3}$ grown crystals \cite{Srikanth97a}. Instead a two-component model
has to be considered as is done below.

\subsection{Comparison to decoupled two-component order parameters}

The simplest way in which one can analyze a two-component system is to
consider two parallel superconducting channels and to add their complex
conductivities. The total conductivity can then be written as:

\[
\sigma =\sigma _1-i\sigma _2=(\sigma _{1A}+\sigma _{1B})-i(\sigma
_{2A}+\sigma _{2B}) 
\]

For ease of calculation we consider that the two components are decoupled
with distinct transition temperatures, $T_{cA}$ and $T_{cB}\,(>T_{cA})\,$.
(We emphasize that in all likelihood the components are coupled, and
furthermore there is only one phase transition at $T_{cB}$, with $T_{cA}$
more of a crossover temperature, as discussed later). Calculations of the
conductivities can be performed using the Mattis-Bardeen formalism
introduced earlier. In an earlier paper, we had presented the qualitative
model where for ease of calculation we used $s-$wave order parameters for
both the $A$ and $B$ components \cite{Srikanth97a}. In reality, it is
essential to choose at least one component to be $d-$wave to produce the
linear low temperature dependence. Using a combination of $s$ and $d$ gap
symmetries, it is possible to obtain excellent quantitative fits to the
observed experimental data for all the $YBCO/BZO$ single crystals.

\begin{figure}[tbph]
\begin{center}
  \includegraphics*[width=0.45\textwidth]{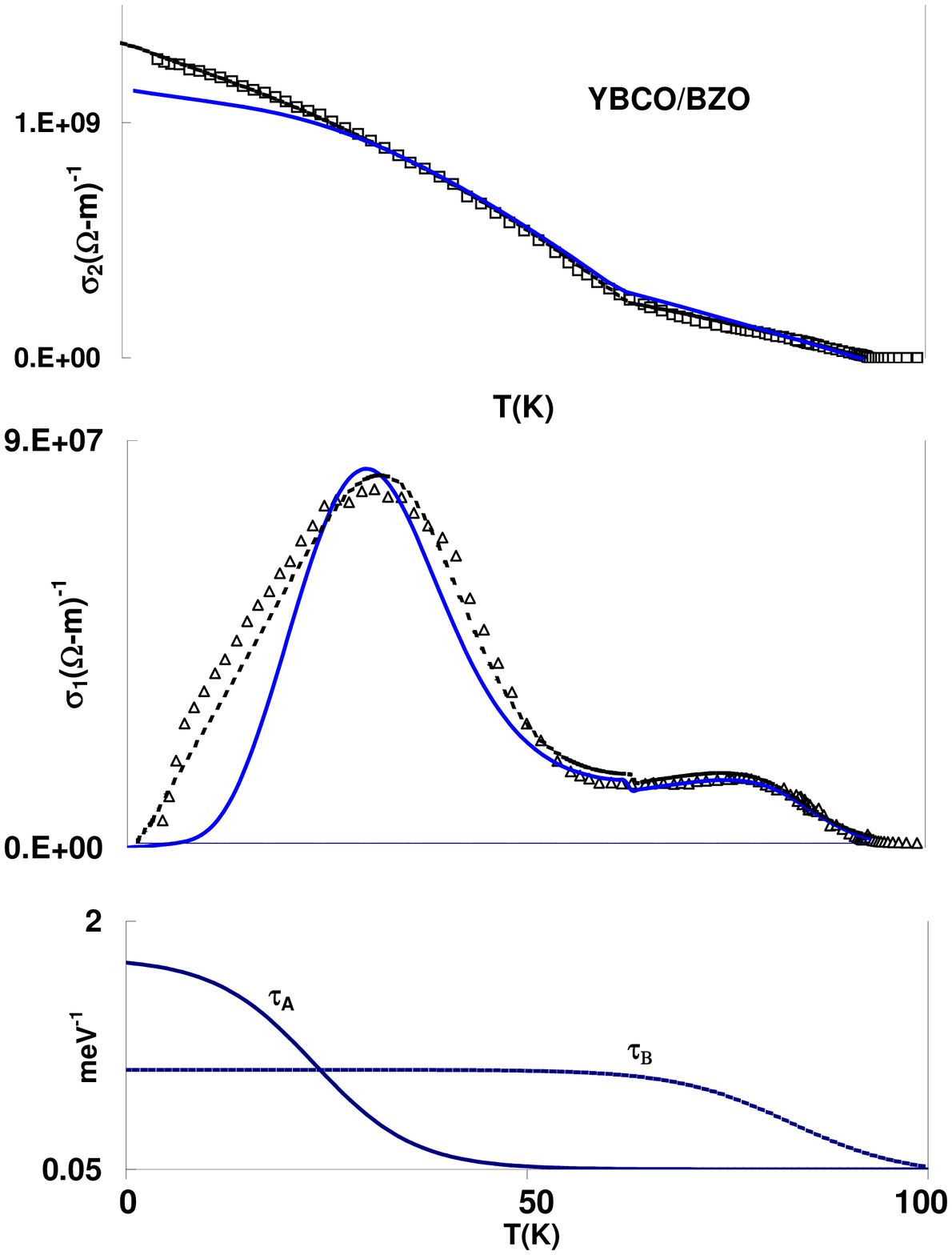} 
\end{center}
\caption{$\sigma _2$ and $\sigma_1$ data for $YBCO/BZO$ single crystal (AE103) along with
the fits generated using a decoupled two-component model (s + d) [solid line] and (d + d) 
(dashed line), discussed in
the text. The bottom panel shows the variation of scattering times $\tau _A$ and $\tau _B$ 
below the $T_c$ for A and B components for the (s+d) case. Essentially similar fits can be obtained for 
(d + s) case (not shown). }
\label{bzofit}
\end{figure}

Fig.\ref{bzofit} shows the experimental data for $%
\sigma _2$ and $\sigma _1$ along with the fits obtained with the two
component model assuming an (s+d) case for the A and B components. Good fits 
can also be obtained for (d+d) and (d+s) cases for the two components. 
The data for the AE103 sample only is presented in the figures for clarity.
It is possible to obtain similar fits for the other $YBCO/BZO$ crystals and
the fit parameters for all the samples are tabulated in Table II.
A good agreement between the data and model is obvious. Best fits are
obtained for the following {\em generic} choices:

\begin{enumerate}
\item  A low temperature component $A$ with $T_{cA}\sim 60K$ and a high
temperature component $B$ with $T_{cB}\sim 93K.$

\item  A weak coupling $d-$ wave gap with $\Delta _{B}=2.16k_{B}T_{c}$ for
the the $B-$ component and an $s-$wave gap with $\Delta _{A}=1.76k_{B}T_{c}$
or a weak-coupling $d-$ wave gap for the $A-$ component.

\item  A lower fraction of $s-$wave ($\sim 0.3)$ than the $d-$wave part ($%
\sim 0.7)$.

\item  Temperature dependent scattering rates $\tau _{A,B}^{-1}\equiv \Gamma
\,_{A,B}=\Gamma \,\,_{A,B(0)}(e^{\alpha (t-1)}-e^{-\alpha })+\Gamma
\,\,_{A,B}^{*}$ were used for for $\sigma _{1}$. This exponential variation
agrees with the suggestion that the scattering time increases rapidly below
the transition temperature. The variation of $\tau _{A}$ and $\tau _{B}$
with temperature is plotted in bottom of Fig.\ref{bzofit}. {\em The detailed
functional form is not as important as the key feature of rapid variation of 
}$\tau _{A}${\em \ below }$65K${\em \ and }$\tau _{B}${\em \ below }$93K$%
{\em .}
\end{enumerate}

\begin{figure}[tbph]
\begin{center}
  \includegraphics*[width=0.45\textwidth]{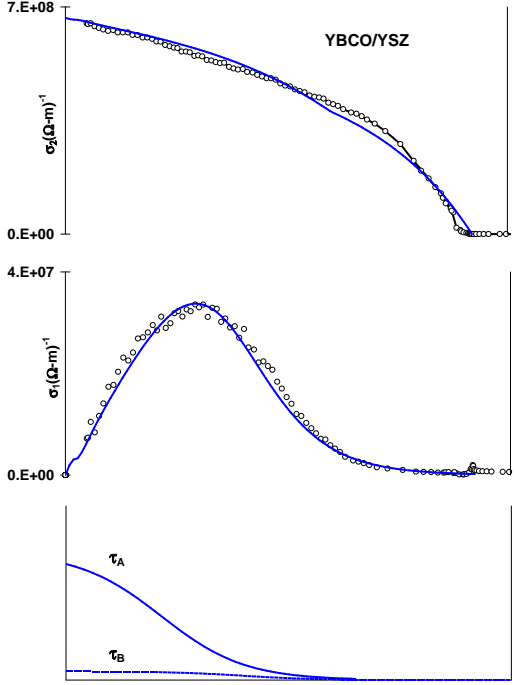} 
\end{center}
\caption{$\sigma _2$ and $\sigma_1$ data for $YBCO/YSZ$ single crystal (AEXX) along with
the fits generated using a decoupled two-component model (s + d) discussed in
the text. The bottom panel shows the variation of scattering times $\tau _A$ and $\tau _B$ 
below the $T_c$ for A and B components for the (s+d) case. The fit is identical with similar 
parameters for the (d + d) case. The (d + s) case does not fit the data. }
\label{yszfit}
\end{figure}

The two-component model also provides an explanation of the $YBCO/YSZ$ data
as can be seen from Fig.\ref{yszfit}. The fit parameters (Table I) suggest that impurities
suppress the order parameter for the A component(a smaller ratio 10\% of the
superfluid density $n_{sA}$ as well as a weaker gap $\Delta
_{A}(0)/k_{B}T_{cA}\sim 1.0$ are needed), as well as greatly enhancing the
scattering rate $\tau _{B}^{-1}$ about 10 times larger for the B component. The rapid
variation of $\tau _{A}$ below $50K$ is still needed to obtain the $30K$
peak in $\sigma _{1}(T)$ for the $YBCO/YSZ$ crystals.

The occurence of the low temperature normal conductivity peak in $\sigma _{1}
$ at around $30K$, even though the superconducting transition temperatures
are in excess of $90K$ in $YBCO$ crystals, has long been a puzzling feature.
An exponential reduction of the quasiparticle scattering rate below $T_{c}$
was invoked to account for this. However, the present data provides an
explanation for the location of this peak, since in the present experimental
results, it is natural to associate the two $\sigma _{1}$ peaks with the
corresponding features in the pair condensate density as inferred from $%
\sigma _{2}$. Two types of pairing with different characteristic energy
scales are clearly observed in these high quality $YBCO/BZO$ single
crystals. In the $YBCO/YSZ$ samples, these energy scales are also present
but their signatures are obscured by impurity scattering.

One aspect of the comparison should be noted. Because we have used a model
in which the components are decoupled, the calculations necessarily show a
sharp break at around $T_{cA} \sim 60K$, whereas the data display a smooth
crossover. This is an indication that the decoupled model is too simplistic,
and a coupled model is necessary. Indeed such calculations 
performed \cite{SSridhar97c} within a Ginzburg-Landau framework lead to
a smooth crossover in $\lambda ^{-2}(T)$ in closer agreement with the
experiments. However a full calculation of the conductivities requires a
microscopic model which still needs to be implemented. (See also the remarks
in the Summary section of this paper).

We note that there have been other discussions of multiple components (gaps
or quasiparticles) in the cuprate superconductors. A two-gap model was
proposed earlier by Klein et al. \cite{NKlein93} to describe the data on $%
YBCO$ thin films with low cation disorder. In their case, however, low and
high gaps both with $s-$wave symmetry and the same $T_{c}$ were chosen to
calculate the pair conductivity and the issue of $\sigma _{1}$ was not
addressed in detail. Two-component behavior in $124$ and underdoped $123$
compounds have also been reported in Raman \cite{Cardona91} and femtosecond
optical response \cite{Mihailovic97}.

Anisotropic penetration depth measurements along the $a$ and $b-$axes
reported on untwinned $YBCO/YSZ$ single crystals \cite{Zhang94}. Lower $%
\lambda (0)$ in the {\it b}-direction suggested superconductivity in the $%
Cu-O$ chains with the onset $T_{c}$ same as that of the planes viz. $93K$.
The similarity between $1/\lambda _{a}^{2}$ and $1/\lambda _{b}^{2}$ was
taken as evidence that chain order parameter also has nodes in the gap.
However, in our present data, additional superfluid response clearly turns
on {\em below} $60K$ and if this is due to chains, then this would imply
that chain superconductivity occurs at a lower temperature than the planes
and would also directly contradict the conclusions reached in ref.\cite
{Zhang94}. Recent experiments on anisotropic thermal conductivity in
detwinned $YBCO/YSZ$ single crystals have also revealed enhanced superfluid
density below $\sim 55K$ which has been interpreted as due to chains \cite
{Gagnon97}. This will be consistent with our observations of the microwave
response in $YBCO/BZO$.

\subsection{Near-Tc behavior:}

We turn our attention now to the transition region close to $T_{c}$. The
temperature dependence of the penetration depth $\lambda (T)$ is expected to
reveal the characteristic nature of the superconducting gap as it opens up
and also possible contributions due to fluctuations. In contrast to
conventional superconductors, the low dimensionality and small coherence
length in cuprates make them good candidates for studying the effect of
superconducting fluctuations and in determining the universality class they
belong to. While Ginzburg-Landau theory generally describes the region near
the transition very well in conventional superconductors, it has been argued
that in $YBCO$ assuming the simplest case of a {\em single} complex order
parameter, it is possible to observe critical behavior which would give rise
to fluctuations corresponding to the universality class of the three
dimensional (3D) $XY$ model. Transport and thermodynamic measurements in the
presence of a magnetic field have been interpreted in terms of this model 
\cite{Salamon93,Overend94}. However, this has also been disputed by Roulin
et al. \cite{Roulin95}{\ }who argue that fits of the specific heat near $%
T_{c}$ in YBCO single crystals for high magnetic fields do not weigh in
favor of 3D $XY$ scaling.

Penetration depth data close to the transition region will be influenced by
the presence or absence of critical fluctuations. Specifically, a plot of [$%
\lambda (0)/\lambda (T)$]$^n\ $as a function of $T$ just below $T_c$ is a
useful indicator of the validity of standard mean-field or the 3D $XY$
models. While a mean-field behavior can be deduced if [$\lambda (0)/\lambda
(T)$]$^2$ is linear in $T$, 3D $XY$ would require [$\lambda (0)/\lambda (T)$]%
$^3$ to be linear in $T$. The latter behavior has been observed in
penetration depth measurements by the UBC group \cite{Kamal94} and has been
interpreted as consistent with the critical behavior of the 3D $XY$ model.
It is important to note the the choice of the range of temperature over
which one looks for effects due to fluctuations is crucial. The dynamical
fluctuations in the frequency dependent microwave conductivity above $T_c$
has been examined by Booth et al. \cite{Booth96}. They arrive at the
conclusion that while the Gaussian fluctuations can account for the behavior
over a larger range in temperature, just $1-2K$ above $T_c$, the behavior is
dominated by critical fluctuations. It should be noted that the temperature
dependence of [$\lambda (0)/\lambda (T)$]$^2$ from kinetic inductance
measurements has shown mean-field dependence \cite{Lemberger96}.

\begin{figure}[tbph]
\begin{center}
  \includegraphics*[width=0.45\textwidth]{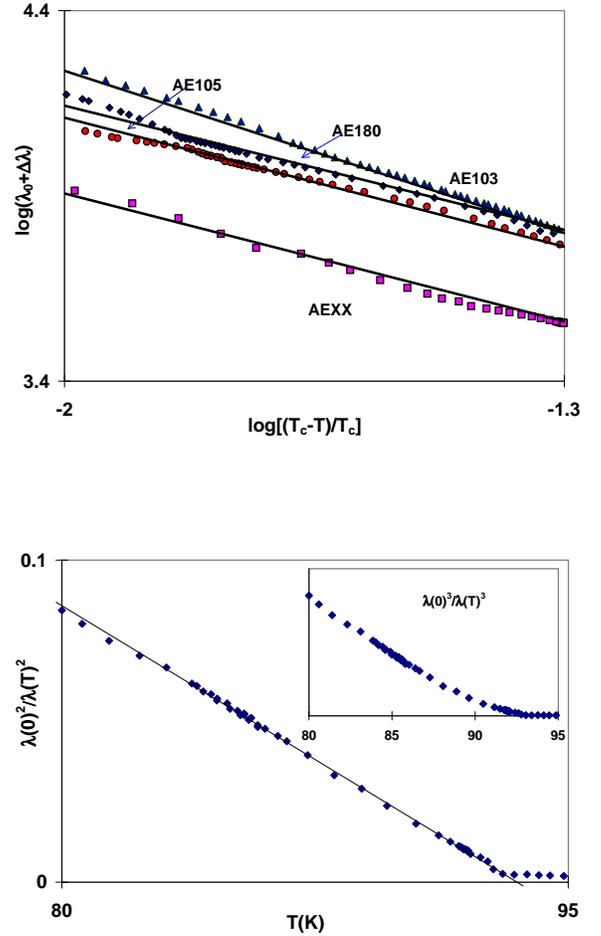} 
\end{center}
\caption{Top panel: Penetration depth near $T_c$ shown in a log-log scale. The data
sets indicate a similar variation in all four samples, with a slope of 0.5
indicating mean field behavior. Bottom panel: Plot of the superfluid density, [$\lambda (0)/\lambda (T)$]$^2$ vs $%
T$ just below $T_c$ for AE103 crystal. The solid line is a guide to the eye
indicating the linear variation consistent with mean-field behavior. Inset
shows [$\lambda (0)/\lambda (T)]^3$ vs $T$ for the same data in which the
curvature is clearly evident indicating disagreement with the 3D XY model.}
\label{meanfield1}
\end{figure}

In a bid to investigate the transition region in our data on $YBCO/BZO$ and $%
YBCO/YSZ$ single crystals, we have plotted the penetration depth for all
four crystals as a log-log plot against the reduced temperature $%
[(T_c-T)/T_c]$ shown in the top panel of Fig.\ref{meanfield1}. It is clear that all the data show a
linear behavior (with a negative slope) with typical values of the slope
lying around $0.5$. The appropriate $\lambda (0)$ and $T_c$ values estimated
from our $R_s$ and $X_s$ data for the three $YBCO/BZO$ samples (AE103,
AE180, AE105) and the $YBCO/YSZ$ sample (AEXX) were taken to obtain the
individual curves. Except the data for AE180 which has a slightly higher
slope ($\sim 0.6$), the other data form a set of parallel lines. This
behavior is consistent with mean-field variation of the order parameter
below $T_c$. To illustrate the point further, a plot of [$\lambda
(0)/\lambda (T)$]$^2$ vs $T$ is presented in the bottom panel of Fig.\ref{meanfield1} for one $%
YBCO/BZO$ crystal. The {\em linear} variation is quite obvious and is
clearly different from the behavior reported for $YBCO$ single crystals in
ref.\cite{Kamal94}. The inset shows the same data plotted as [$\lambda
(0)/\lambda (T)$]$^3$ vs $T$ which is non-linear and rules out the
possibility of 3D $XY$. The smooth variation of the [$\lambda (0)/\lambda
(T) $]$^3$ vs $T$ curve also indicates against the possibility of
interpretation as a crossover behavior from 3D $XY$ very near $T_c$ to a
mean-field like variation at lower temperatures. Our conclusion in this
context would be that the near-$T_c$ behavior in all $YBCO$ crystals is
governed by standard mean-field expression. It is interesting to note that
this universal behavior may also be indicative of a scenario against the
possibility of a single order parameter. Absence of 3D $XY$ correlations is
not surprising if one considers unconventional pairing states with mixed
symmetry.

\subsection{Normal state scattering rate}

Assuming a local $j-\hat{E}$ relation in the skin depth limit, the normal
state surface resistance can be written as $R_{n}=\sqrt{\omega \mu _{0}\rho
_{n}/2\text{ }}$ where the microwave normal state resistivity is expected to
be the same as the DC resistivity and $\rho _{n}=2\Gamma \,\,\mu _{0}\lambda
(0)^{2}.$ Here $\lambda (0)$ is the London penetration depth and $\Gamma
\,\, $ is the normal state scattering rate. Taking $\rho _{n}=\rho
_{0}+\gamma T$ the resistivity values can be translated into the inelastic
scattering rates given by $\Gamma \,\,=\gamma T/2\mu _{0}\lambda ^{2}(0).$

\begin{figure}[tbph]
\begin{center}
  \includegraphics*[width=0.45\textwidth]{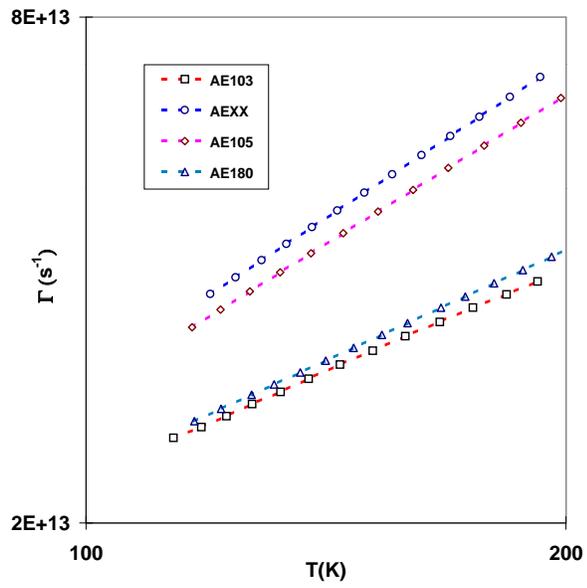} 
\end{center}
\caption{The normal state inelastic scattering rate for all four samples.
The data for AE103 and AE180 have lower values and smaller slope in
comparison to AEXX and AE105 samples. }
\label{inelastic}
\end{figure}

In Fig.\ref{inelastic}, we have plotted the normal state inelastic
scattering rate as a function of temperature for $100K<T<200K$ for all the $%
YBCO/BZO$ samples along with the data for $YBCO/YSZ$ crystal. Among the $%
YBCO/BZO$ single crystals, the $\Gamma $ values as well as the slopes for
AE103 and AE180 over the entire temperature range are lower than that for
the AE105 sample. The $YBCO/YSZ$ AEXX sample has the largest $\Gamma $ in
this batch of crystals whose microwave surface impedance were measured in an
identical manner. A comparison of this data with the $\sigma _1(T)$ data of
Fig.\ref{sigmas} shows a clear correlation between the normal state
scattering rate and the new $80K$ quasiparticle conductivity peak. This peak
is prominent only in AE180 and AE103 samples which have a lower scattering
rate, is suppressed in AE105 which shows a larger $\Gamma $ than the other
two $YBCO/BZO$ samples and practically absent in AEXX which has the largest
scattering rate. This remarkable correlation indicates the sensitive nature
of the new conductivity peak to very small changes in impurity concentration
and/or oxygen stoichiometry. Further studies in $YBCO/BZO$ crystals with
transition-metal doping and variation of oxygen stoichiometry are needed to
explore the correlation between the electronic properties in the normal
state and the quasiparticle contribution in the superconducting state which
is suggested by our present data.

\section{Summary of Microwave properties of YBCO/BZO and YBCO/YSZ crystals
}

We summarize below some of the microwave properties exhibited by the $%
YBCO/BZO$ crystals along with comparison of results on the previous class of 
$YBCO/YSZ$ samples. The key observations are:

\begin{enumerate}
\item  The linear behavior of the low temperature penetration depth $\lambda
(T)\propto T$ appears to be very robust and is observed in all samples with
essentially the same slope. $\lambda (T)$ rises linearly with $T$ upto $%
0.25T_{c}$ in $YBCO/BZO$ single crystals with a characteristic slope ($%
d\lambda /dT)$ of about $5\AA /K$. This linear variation is in agreement
with the observations on $YBCO/YSZ$ crystals too and has been ascribed to
the presence of nodes in the gap, and overall is considered a strong
evidence for $d-$wave order parameter \cite{Hardy93a}. However, unlike the
data in $YBCO/YSZ$, the full temperature dependence for $YBCO/BZO$ display a
non-monotonic behavior which points to the presence of an extra component in
addition to a $d-$wave part.

A linear behavior of $\lambda (T)$ is also seen in good quality $BSCCO$
single crystals \cite{TJacobs95d,SFLee96,Shibauchi96}. A $T^{2}$ behavior
which has also been reported in $YBCO$ thin films \cite{Ma93}, has been
considered a consequence of inhomogeneties and additional scattering caused
by impurities and disorder.

\item  The surface resistance $R_{s}(T)$ of $YBCO/BZO$ shows a bump at
around $30K$ as well as a broad bump at higher temperatures. In terms of
conductivity, this results in two peaks occuring at $\sim 30K$ and $\sim 80K$%
. The ratio $\sigma _{1}/\sigma _{n}$ of the peaks are much higher and
cannot be explained due to coherence effects predicted by BCS theory. We
have associated these two peaks with two different quasiparticle systems
corresponding to components A and B as discussed earlier.

In $YBCO/YSZ$, only the low temperature peak at $\sim 30K$ is seen. This was
first observed at sub-THz \cite{Nuss91} and at microwave frequencies \cite
{Bonn92a}. To account for this feature, Bonn et al. \cite{Bonn92a} proposed
an rapid increase in the inelastic scattering time $(\tau )$ below $T_{c}$
followed by a reduction in carrier density at low temperatures.

\item  The interpretation of the conductivity peak as arising from an
interplay of increasing $\tau $ and decreasing $n_{qp}$ as $T$ is lowered
has also been used here to account for the conductivity peak \cite
{Nuss91,Bonn92a}. Such an interpretation necessarily requires a rapid
variation of $\tau _{B}$ just below $93K$ and of $\tau _{A}$ below $65K$ as
is evident from Fig. \ref{bzofit} in $YBCO/BZO$.

It is to be noted that in $BSCCO$, our results \cite{TJacobs95d} do not show
any conductivity peak in $\sigma _{1}$ below $T_{c}$ but rather $\sigma
_{1}/\sigma _{n}$ rises monotonically as $T$ is lowered. However another
study \cite{SFLee96}claims to see a small reduction in $\sigma _{1}/\sigma
_{n}$ at very low temperatures suggestive of a conductivity peak.

\item   It is important to note that {\em the two temperature scales of }$93K
${\em \ and }$65K${\em \ are present in }$YBCO/YSZ${\em \ also}. This can be
seen from the fact that two component analysis also describes the $YBCO/YSZ$
data, as shown in Fig.\ref{yszfit}, provided a suppressed $n_{sA}$ and $\tau
_{B}$ are used. This suggests that in the $YBCO/YSZ$ crystals, impurities
lead to suppression of the pairing density of the A component, as well as
greatly enhanced scattering of the B component.

\item  A very sharp peak in $\sigma _{1}$ just below $T_{c}$ at around $%
91-92K$ is seen in all ($YBCO/BZO$ as well as $YBCO/YSZ$ ) samples, so that
there are a total of $3$ conductivity peaks. This peak near $T_{c}$ has been
ascribed to fluctuations \cite{vanSarloos91} or inhomogeneities \cite
{Olsen91}. In the $YBCO/BZO$ samples this peak is extremely sharp and
further attests to the high sample quality.

\item  The temperature scale of $65K$ naturally raises the possibility of a
multi-phase sample with regions of $YBCO_{6.5}$. We have considered this
issue very carefully and several experiments show that this is {\em %
conclusively} ruled out. In-situ resisitivity meaasurements at high
temperatures have shown that the crystals can be reversibly
oxygenated-deoxygenated, and that the diffusion constants are well
characterized, indicating nothing unusual about the oxygenation of these
samples \cite{Erb96a}. Note also that the $YBCO/YSZ$ sample $AEXX$ was
oxygenated using the same procedures but did not display the new features in
the conductivity. Furthermore, the similarity of the AE180 data with the
other samples conclusively shows that even local $O$ vacancies are not
responsible for the present results. This is because in AE180, high pressure
oxygenation breaks up $O$ deficient clusters, as evidence by the elimination
of the fishtail anomaly in samples prepared by this method.

\item  Our measurements on both $YBCO/BZO$ and $YBCO/YSZ$ single crystals
are consistent with mean-field behavior rather than $3DXY$ close to $T_{c}$.
Near-$T_{c}$ variation of $1/\lambda ^{2}$ has been analysed in terms of $%
3DXY$ scaling by Kamal et al. \cite{Kamal94} in their $YBCO/YSZ$ crystals.
While a more rigorous study of the fluctuation contributions below and above 
$T_{c}$ by Booth et al. \cite{Booth96} indicated that $3DXY$ scaling is seen
only over a limited temperature range, kinetic inductance measurements by
Lemberger et al. \cite{Lemberger96} on thin films suggest mean-field
behavior, in agreement with the conclusions of this paper.

\item  It should be noted that the pairing onset below $65K$ in the $\sigma
_{2}(T)$ is quite broad, and indicates that this is not likely to be a
critical point. {\em Thus we believe that there is only one transition at }$%
T_{c}=93.4K${\em , with a crossover at }$65K${\em \ where the additional
pairing channel opens up. }Furthermore, although we have used a decoupled
scenario to analyze the data, it is very likely that the OP's are coupled.
This is confirmed by calculations of the superfluid density using a simple
Ginzburg-Landau free energy for two coupled OP, which reproduce the main
features of the $\lambda ^{-2}(T)$ data \cite{SSridhar97c}. At high
temperatures $T_{c}>T>T_{cA}$, the symmetry is predominantly of type $B$,
although a small $A-$component is also induced due to the coupling. At lower
temperatures $T<T_{cA}$, the $A-$component grows so that both components
(and hence a mixed symmetry) are present.  Microscopic calcuations of
thermodynamic properties of superconducting lattice fermions which also have
multiple pairing channels also show that the thermodynamic signatures, such
as specific heat cusps, are weaker in the subdominant channels which appear
at lower temperatures than in the main channel \cite{Otnes97}.

While we have presented a simple phenomenological two-component model in
this paper to account for the two conductivity peaks in $YBCO/BZO$,
microscopic calculations (similar to those based on a single $d-$wave order
parameter \cite{Klemm88,Hirschfeld94,Scharnberg97}) need to be done for the
surface impedance of a two-component superconductor. Furthermore, the
analysis used here assumes phase coherence throughout the sample and does
not include contributions arising from phase fluctuations\cite
{Emery95,Stroud95}.
\end{enumerate}

\section{Relation to Microscopic theories}

Models based on any type of mixed order parameter symmetry would yield
results consistent with our experimental observations in the $YBCO/BZO$ system.
Mixed order parameter symmetry and multi-component behavior of the order
parameter has been addressed in several theoretical papers which can be
broadly classified along the following categories:

(1){\it \ }${\em s+d,s+id:}$ It has been argued that the orthorhombic
structure of $YBCO$ naturally leads to a mixing of $s$ and $d$ order
parameter components \cite{Koltenbah97}. Several theories \cite
{Muller95,Betouras95,Odonovan95,Maki96,Otnes97,vanderMarel95a} advocate to
this general concept of a mixed $s+d$ OP and have successfully developed
models which can describe the experimental results in $YBCO.$ This appears to be 
consistent with photoemission data on overdoped $BSCCO$ also \cite
{Onellion95,Betouras95}. Phenomenological \cite{Betouras95} and microscopic
theories \cite{Ren96} have been considered. The work of ref. \cite{Ren96}
indicates that the system likely goes from a $d-$wave state at high
temperatures to an $s+id$ state at low $T$.

(2){\em Chain-Plane coupling:} The presence of $Cu-O$ chains in addition to
planes raises the possibility of different superconducting condensates with
different pairing energies residing on chains and planes. This can
effectively lead to a two-component system exhibiting two different gaps 
\cite{Kresin92}. Coupling between chain and plane bands has been proposed 
\cite{Combescot95} to account for both the $s$ and $d$ characteristics
displayed by $YBCO$.

(3){\em Multi band models and Fermi pockets:} Theoretical treatments based
on multi band models have been presented which considers the interband
interactions \cite{Doniach93,Combescot95,Combescot97,Golubov96,Taraphder97}.
This general framework also reproduces the essential features of a
two-component superconductor. The nature of the Fermi surface could be
important in determining the possibility of multiple pairing energies. The
presence of pockets with $d$ and $g-$ wave symmetries has been recently
discussed \cite{Sushkov97}.

(4) {\em Surface states and Time Reversal Symmetry breaking:} The
possibility of mixed order parameter symmetry at the surface leading to
breaking time reversal symmetry has been suggested \cite
{Laughlin94,Tikofsky95}. Recent observation of Andreev bound states at the
surface of $YBCO$ \cite{Greene97,Sauls97} strongly suggests this possibility.

(5) {\em Interlayer Tunneling:} In a recent paper, Xiang and Wheatley \cite
{Xiang96} presented a model based on proximity effect incorporating a
microscopic pair tunneling process which couples the $Cu-O$ chains and
planes in $YBCO$ , to account for the experimental results on $YBCO/YSZ$
data from the UBC group \cite{Zhang94}. Our present results on the $YBCO/BZO$
crystals suggests that within the interlayer tunneling model, single
electron tunneling processes could account for the features observed in our
data.

\section{Conclusion}

In summary, we have shown that high quality single crystals of $YBCO$ grown
in $BaZrO_{3}$ crucibles exhibit features in their microwave properties,
which are inconsistent with a pure d-wave OP and instead point to the
occurence of multi-component superconductivity.The measurements reveal the
presence of two pairing temperatures corresponding to two superconducting
(pair/quasiparticle) components in the optimally doped compound. This should
be borne in mind and careful studies should be performed on high quality
materials to further explore these issues.

\acknowledgements

Work at Northeastern was supported by NSF-DMR-9623720, and at Geneva by the
Fonds National Suisse de la Recherche Scientifique. We thank R. S.
Markiewicz, Balam A. Willemsen and D. P. Choudhury for useful discussions.

\end{multicols}

\widetext
\begin{table}
\label{table.1}
\begin{tabular}{l|l}
\textbf{YBCO/YSZ} & \textbf{YBCO/BZO} \\ \hline
Crucible: Yttria-stabilized Zirconia (YSZ) & Crucible: $BaZrO_3$ \\ \hline
Crystal Purity: 99.5 to 99.95 \% \cite{Casalta96} & Crystal Purity: 99.995
\% \cite{Erb96a} \\ \hline
$T_c=93.2$ to $93.5K$ & $T_c=93.2$ to $93.5K$ \\ \hline
Estimated $\lambda (0)\sim 1600-2000\AA $ & Estimated $\lambda (0)\sim
1000-1400\AA $ \\ \hline
Flux lattice not imaged with STM &   Flux lattice imaged with cryogenic STM 
\cite{MaggioAprile95} \\ \hline
Schottky anomaly present in specific heat &  Schottky anomaly
suppressed in high pressure \cr & oxygenated  $O_{6.95}$ and eliminated in $O_7$ 
\cite{Roulin96} \\ \hline
Large ''Fishtail effect'' in Magnetization & ''Fishtail'' greatly suppressed
in high pressure \cr & oxygenated $O_{6.95}$ and eliminated in $O_7$ \cite{Erb3}
\\ \hline
Substantial Pinning due to impurities & Very low pinning evidenced by
extremely low critical \cr & currents and observation of flux lattice melting \cite
{Indenbom97} \\
\end{tabular} \vspace{20mm}
\caption{A comparison of some material and physical properties of $YBCO/BZO$
and $YBCO/YSZ$ single crystals.}
\end{table}
\break

\widetext
\begin{table}
\label{table.2}
\begin{tabular}{c|c|c|c}
\textbf{Samples} & \textbf{AE103 (\it d + \it d)} & \textbf{AE105 (\it d + \it d)} & \textbf{AE180 (\it d + \it d)} \\ \hline
\textbf{Components} & \textbf{A(0.6), B(0.4)} & \textbf{A(0.6), B(0.4)} & \textbf{A(0.48), B(0.52)} \\ \hline
$T_{c(A,B)}$ & 63K, 93K & 59K, 93K & 58K, 93K \\ \hline
$\Gamma_{(A,B)(0)}$ meV & 11, 8 & 10, 13 & 7, 13 \\ \hline
$\alpha$ & 12, 25 & 10, 8 & 9, 14 \\ \hline
$\Gamma^*_{(A,B)}$ meV & 0.08, 0.6 & 0.08, 1.1 & 0.06, 1.2 \\
\end{tabular}
\vspace{10mm}
\begin{tabular}{c|c|c|c}
\textbf{Samples} & \textbf{AE103 (\it s + \it d)} & \textbf{AE105 (\it s + \it d)} & \textbf{AE180 (\it s + \it d)} \\ \hline
\textbf{Components} & \textbf{A(0.6), B(0.4)} & \textbf{A(0.6), B(0.4)} & \textbf{A(0.45), B(0.55)} \\ \hline
$T_{c(A,B)}$ & 63K, 93K & 59K, 93K & 58K, 93K \\ \hline
$\Gamma_{(A,B)(0)}$ meV & 12, 8 & 18, 14 & 10, 16 \\ \hline
$\alpha$ & 11, 25 & 11, 6 & 10, 16 \\ \hline
$\Gamma^*_{(A,B)}$ meV & 0.05, 0.8 & 0.04, 0.4 & 0.02, 1.4 \\
\end{tabular}

\vspace{10mm}
\begin{tabular}{c|c|c|c}
\textbf{Samples} & \textbf{AE103 (\it d + \it s)} & \textbf{AE105 (\it d + \it s)} & \textbf{AE180 (\it d + \it s)} \\ \hline
\textbf{Components} & \textbf{A(0.78), B(0.22)} & \textbf{A(0.78), B(0.22)} & \textbf{A(0.68), B(0.32)} \\ \hline
$T_{c(A,B)}$ & 63K, 93K & 59K, 93K & 58K, 93K \\ \hline
$\Gamma_{(A,B)(0)}$ meV & 20, 7 & 12, 9 & 10, 10 \\ \hline
$\alpha$ & 11, 21 & 11, 10 & 9, 14 \\ \hline
$\Gamma^*_{(A,B)}$ meV & 0.24, 0.52 & 0.06, 0.01 & 0.1, 0.4 \\
\end{tabular}

\vspace{20mm}
\caption{Fit parameters for the three $YBCO/BZO$ crystals using the several two-component scenarios. For the $YBCO/YSZ$ case, fits shown 
in Fig. \ref{yszfit} using the two-component model are generated using the following parameters: 
Components: A(0.1) and B(0.9); $T_c{A,B} = 60, 92K$; $\Gamma_{(A,B)(0)} = 10, 14$; $\alpha = 7, 12$; $\Gamma^*_{(A,B)} = 0.1, 0.01$. While 
the numbers are identical for (s + d) and (d + d) cases, (d + s) does not fit the data. }
\end{table}
\break

\begin{multicols}{2}
\narrowtext

\break

\end{multicols}
\end{document}